%% file: main.tex
\documentclass[sigconf]{acmart}

\AtBeginDocument{%
  }

\copyrightyear{2025} 
\acmYear{2025} 
\setcopyright{acmlicensed}\acmConference[SIGCSE TS 2025]{Proceedings of the 56th ACM Technical Symposium on Computer Science Education V. 1}{February 26-March 1, 2025}{Pittsburgh, PA, USA}
\acmBooktitle{Proceedings of the 56th ACM Technical Symposium on Computer Science Education V. 1 (SIGCSE TS 2025), February 26-March 1, 2025, Pittsburgh, PA, USA}
\acmDOI{10.1145/3641554.3701940}
\acmISBN{979-8-4007-0531-1/25/02}

\begin{document}

\title[LLM-Feedback for Conceptual Design]{LLM-Driven Feedback for Enhancing Conceptual Design Learning in Database Systems Courses}

\author{Sara Riazi}
\email{riazi@uic.edu}
\affiliation{%
  \institution{Department of Computer Science\\University of Illinois Chicago}
  \city{Chicago}
  \state{IL}
  \country{USA}
}



\author{Pedram Rooshenas}
\email{pedram@uic.edu}
\affiliation{%
  \institution{Department of Computer Science\\University of Illinois Chicago}
  \city{Chicago}
  \state{IL}
  \country{USA}
}


\renewcommand{\shortauthors}{Sara Riazi and Pedram Rooshenas}


\begin{abstract}
The integration of LLM-generated feedback into educational settings has shown promise in enhancing student learning outcomes. This paper presents a novel LLM-driven system that provides targeted feedback for conceptual designs in a Database Systems course. The system converts student-created entity-relationship diagrams (ERDs) into JSON format, allows the student to prune the diagram by isolating a relationship, extracts relevant requirements for the selected relationship, and utilizes a large language model (LLM) to generate detailed feedback. Additionally, the system creates a tailored set of questions and answers to further aid student understanding. Our pilot implementation in a Database System course demonstrates effective feedback generation that helped the students improve their design skills.

\end{abstract}






\maketitle

\input{intro}

\input{bg}
\input{method}

\input{qualitative}
\input{study}

\input{discussion}

\clearpage

\bibliographystyle{ACM-Reference-Format}
\balance
\bibliography{all}


\end{document}

%% file: intro.tex
\section{Introduction}
Conceptual design is a critical component of database systems education, requiring the translation of real-world requirements into formalized database schemas. This process necessitates high-level abstract thinking and a robust understanding of entities and relationships, often posing significant challenges for students. Traditional teaching methods, supplemented by limited instructor and teaching assistant feedback, frequently fail to provide the comprehensive and timely feedback needed to master these skills. Moreover, providing students with more design practice requires even more resources, further straining educational support systems.

Efforts to create virtual learning environments for conceptual design have deep roots, dating back to early systems such as \citet{hall&gordon98}. 
These pioneers provided interactive platforms that allowed students to practice and refine their design skills, addressing limitations inherent in traditional classroom instruction. Modern intelligent tutoring systems (ITS) employ constraint-based models (CBMs) to generate feedback~\cite{zakharov2004feedback,mitrovic2006constraint,suraweera2002kermit}. These systems rely on naming conventions to construct constraints based on the structural and relational integrity of models. Recent advancements have introduced sophisticated methods for semantic matching to mitigate issues related to naming conventions. However, despite these advancements, the feedback generated by these systems often focuses merely on pointing out mistakes related to constraint violations, without a deep contextual understanding of the student's modeling approach. This sometimes limits the systems' ability to offer constructive advice on how to rectify specific modeling errors.


With advances in large language models (LLMs) and their capability to understand sophisticated natural language and generate contextually appropriate answers, new efforts have been initiated to employ LLMs for automatic feedback generation. These models promise to enhance the adaptability and effectiveness of feedback by leveraging their extensive training on diverse datasets.

This work introduces an AI-powered system that leverages LLMs to provide comprehensive feedback on conceptual design tasks in database systems courses. The system converts student-created entity-relationship diagrams (ERDs) into JSON format, allows the student to prune the diagram by isolating a relationship, extracts relevant requirements, and uses an LLM to generate detailed feedback. This feedback is tailored to address specific aspects of the design, such as the identification of entities, relationships, and cardinalities. We study different aspects of feedback generation and evaluate the quality of the generated feedback through student evaluations and expert assessments in a pilot study for a Database Systems course.

%% file: bg.tex
\section{Related Works}
\noindent\textbf{Intelligent Tutoring Systems}: Dialog-based Intelligent Tutoring Systems (ITS) such as AutoTutor~\cite{graesser1999autotutor,graesser2004autotutor} and its successors~\cite{rus2015deeptutor,rus2013recent,d2012gaze,olney2010tutorial,graesser2003autotutor,sullins2015exploring,nye2014autotutor}, have thoroughly explored and shown the efficacy of such systems in teaching subjects like physics and mathematics. These ITS implementations offer step-by-step feedback during problem-solving, leveraging a model of knowledge such as constraint-based models~\cite{mitrovic2007intelligent} or Bayesian networks~\cite{bntutor} to track the state of the student's knowledge relative to the learning objective and generate appropriate feedback. (See \cite{paladines2020systematic} for a systematic review of ITS.) However, the complexity of designing such a knowledge model for each course has hindered the wide adoption of such systems in practice.\\

\noindent\textbf{Feedback generation for diagrams}: 
Previous approaches to automatic feedback generation on diagrams primarily rely on comparing the candidate diagram to a correct solution~\cite{correia&al17,foss&al22}. However, correct diagrams are not unique for a given set of requirements, and students may design their diagrams differently. The existence of multiple correct diagrams makes comparison-based feedback less effective. Additionally, we observe that when the LLM has access to a correct diagram, it tends to make fewer mistakes but often assumes any deviation from the given solution is an error, even in the naming.


In this project, we rely on LLMs for comprehending the JSON representation of the diagrams. We leave alternative approaches such as representing the diagrams as knowledge graphs and use that to guide LLMs~\cite{pan2023unifying} for future studies.\\

\noindent\textbf{LLMs and Education}: Prior research indicates that LLMs can serve as effective tools for generating educational content~\cite{leiker2023prototyping} and high-quality and effective feedback~\cite{feedback_chatgpt, nguyen2024using,balse2023investigating,nguyen2024comparing,liu2024beyond}. Sarsa et al.~\cite{sarsa2022automatic} show that OpenAI Codex -- a commercial large-language model specialized for code generation -- can be used for constructing novel and sensible programming exercises as well as their answers and explanation for educational purposes. Ochieng~\cite{ochieng2023large} shows LLMs can be used for crafting substantial in-context questions that foster guided learning. With appropriate fine-tuning on curated data, LLMs can also be leveraged to automatically assess students' responses to short answer questions~\cite{moore2022assessing}. Finally Khanmingo~\cite{khanmigo}, is a LLM (based on GPT-4) fine-tune for tutoring by engaging the user via asking questions and creating dialog rather than generating the the whole response directly. 


%% file: method.tex
\begin{figure*}
    \centering
    \includegraphics[width=0.8\linewidth]{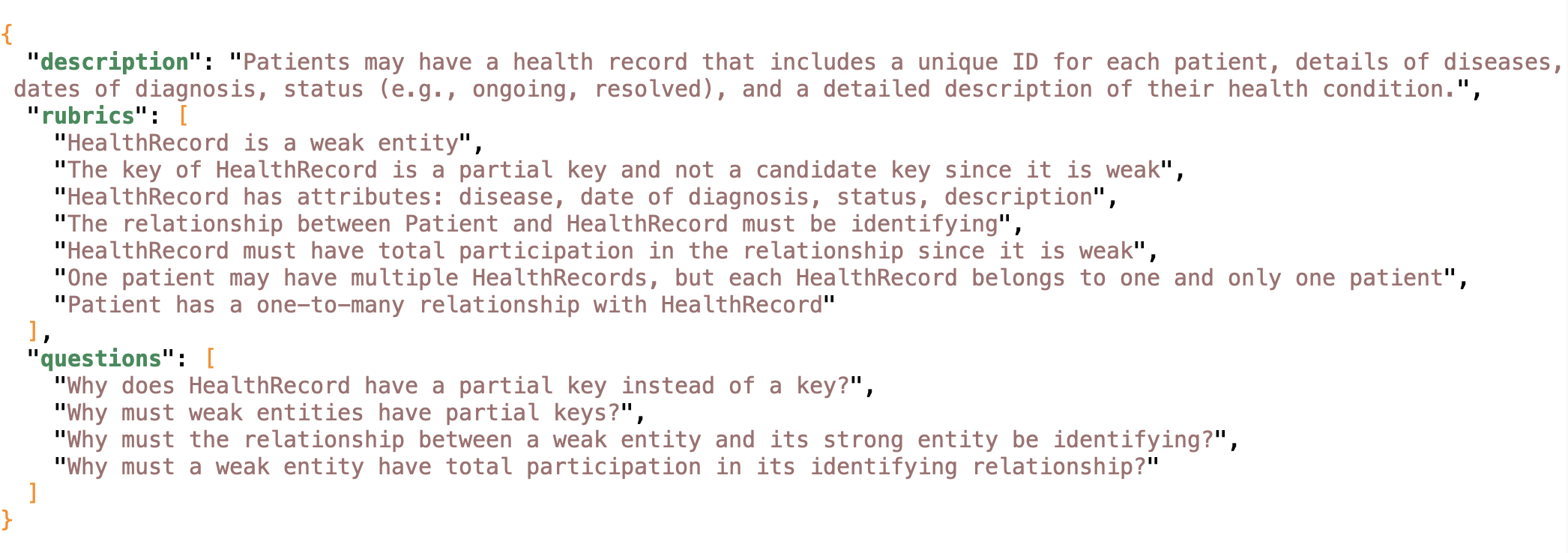}
    \caption{JSON representation of problems statements, including requirements, rubrics, and questions.}
    \label{fig:requirements}
\end{figure*}

\begin{figure}
    \centering
    \includegraphics[width=0.8\linewidth]{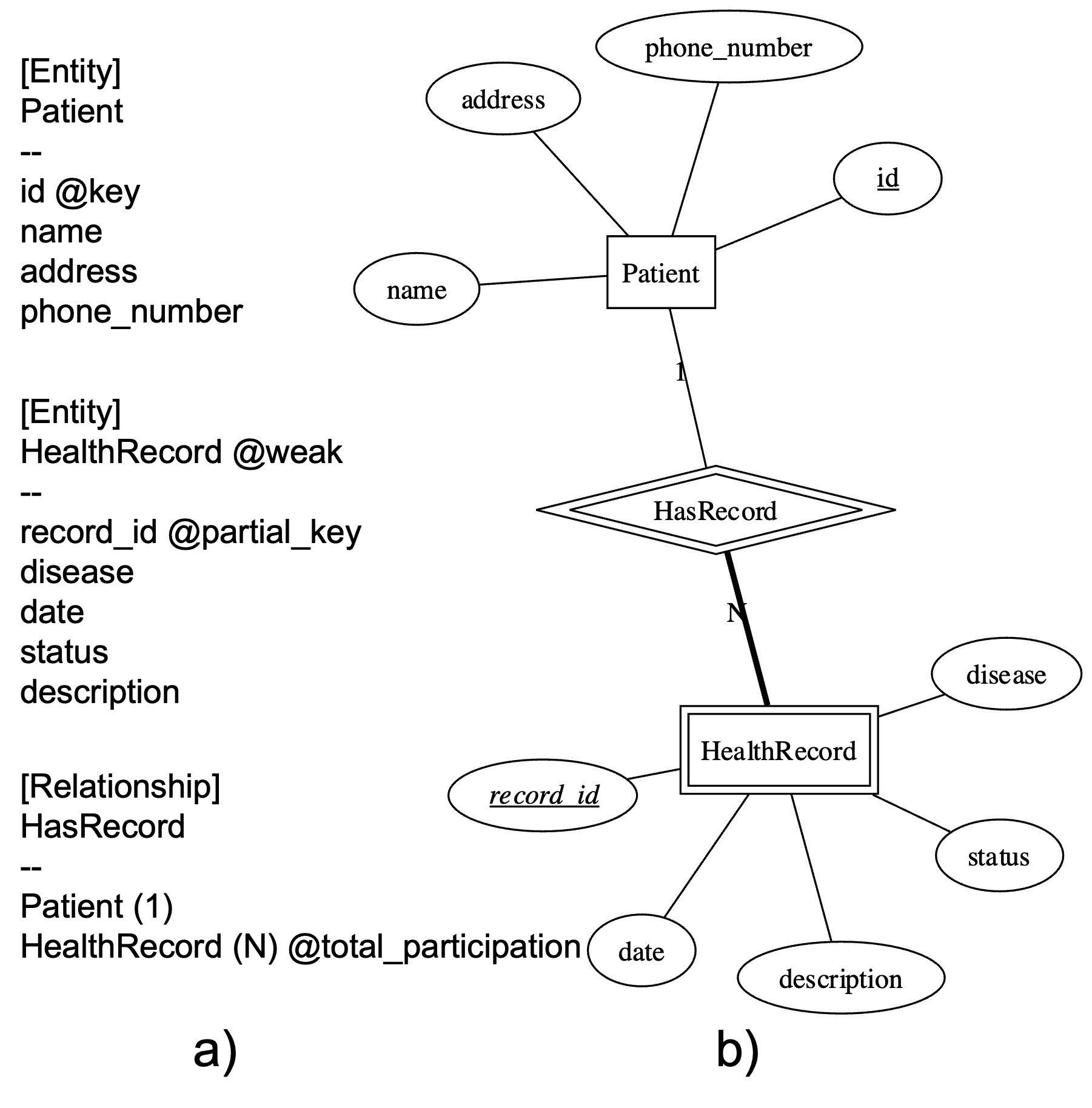}
    \caption{a) Used ERD-notation b) Graphviz visualization of an isolated relationship. }
    \label{fig:erd}
\end{figure}

\begin{figure}
    \centering
    \includegraphics[width=0.8\linewidth]{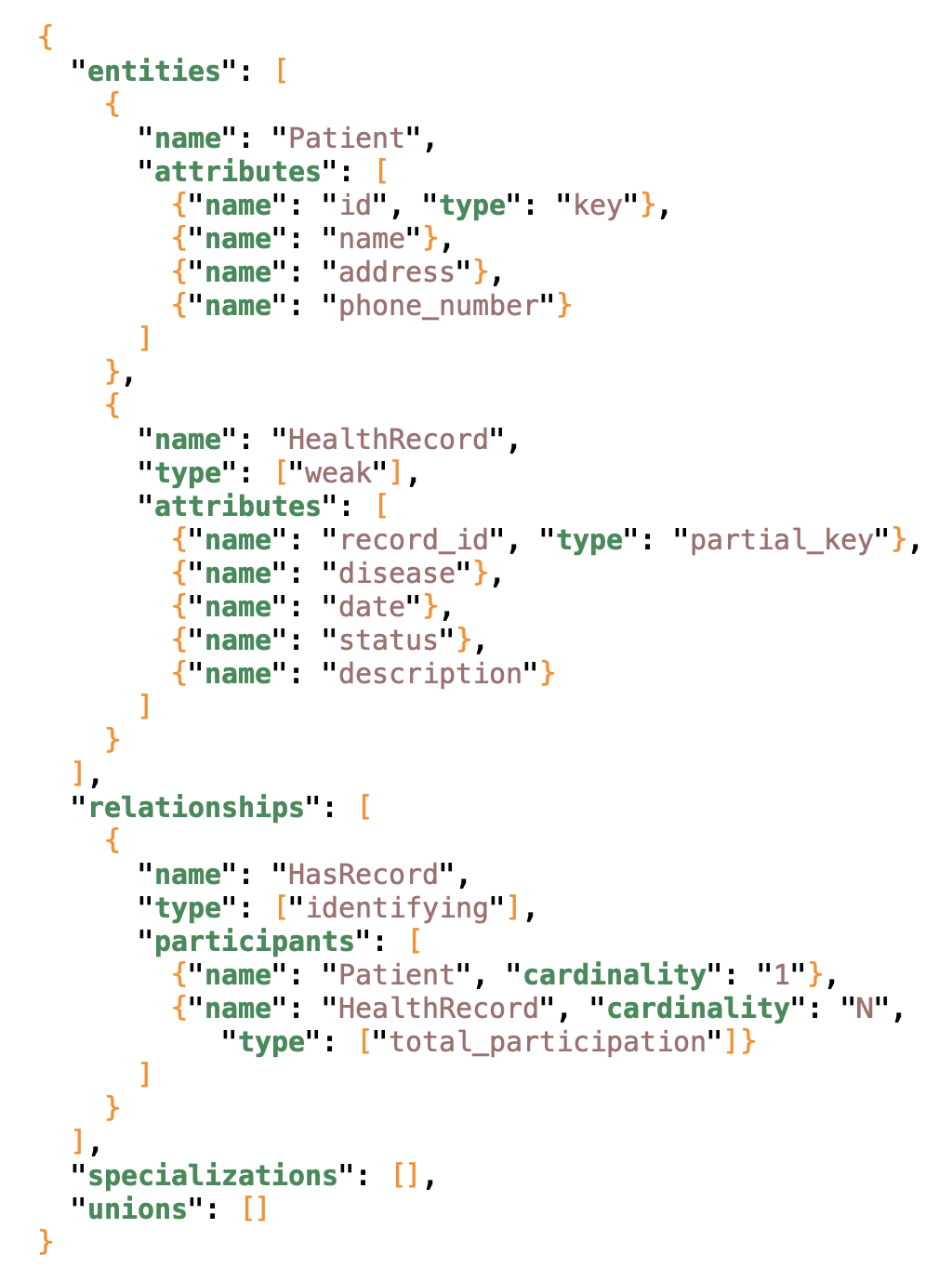}
    \caption{JSON representation of an isolated relationship.}
    \label{fig:erd_json}
\end{figure}

\section{Methodology}
Conceptual design involves mapping a set of requirements into (extended) entity-relationship diagrams ((E)ERDs). To evaluate the consistency of an EERD with the provided requirements, LLMs must be supplied with both the requirements and the diagram as the prompt. By exploiting the language understanding capabilities of LLMs, we can define the constraints using natural language for the requirements rather than for the model, thereby removing the dependency of the constraints on naming conventions. However, we rely on the LLMs to evaluate these natural language constraints, which we refer to as rubrics.

Now, the main question is how to construct the appropriate prompt to receive comprehensive and informative feedback from the LLM.

\subsection{Requirement Representation}
We use JSON -- which is a common representation for structured information and most of the current off-the-shelf LLMs can understand the described -- to represent the requirements instead of plain text. Each piece of requirement consists of three sections: description, rubric, and questions (see Figure~\ref{fig:requirements}). 
The description is what students see and the main problem statements, while the rubric and questions are provided by the educator to guide the LLMs. Although LLMs, especially commercial ones like GPT-4, have been trained on vast amounts of data and possess impressive general knowledge, they may still make mistakes in assessing the diagrams based solely on the descriptions without access to the set of rubrics (see Section for qualitative comparison). We also include a set of possible questions for each requirement item. These questions are carefully designed by the educator. The LLMs use these questions to generate a set of FAQs for each feedback, providing students with more information.

Augmenting the problem description with a set of rubrics and questions is fairly easy for educators, making the system tractable and practical for wide adoption.


\subsection{EERD Representation}
In order to use LLMs to generate feedback on conceptual design, we need to represent entity-relationship diagrams in a text format.

Students submit their ERDs in a text-based grammar as shown in Figure~\ref{fig:erd}.a. After entering the ERD notation, the system translates the entered ERD into JSON format (Figure~\ref{fig:erd_json}). We use Graphviz, an automated graph visualization tool, to visualize the entered ERDs. Students can view the entire ERD or different parts of their ERDs by selecting a relationship (see Figure~\ref{fig:erd}.b). By selecting a relationship, the system prunes the ERD to involve only the relationship and its participating entities. We use the pruned ERD to generate feedback because LLM feedback is more detailed and informative when it comments on specific parts of the diagram rather than the whole.

Providing feedback on a relationship one at a time helps students work on the ERD gradually and carefully follow the requirements covering the relationship and corresponding entities. However, we also found that focusing on a single relationship and its entities can sometimes lead to incomplete feedback as the AI expects more contextual information. For example, if two entities participate in different relationships, the LLM expects both relationships to be present. Similarly, if an entity also participates in a specialization/generalization or union, the LLM may expect those details to be included.

To address this, we extend our pruning process to include such related relationships, ensuring that the LLM has sufficient context to provide accurate feedback.

The JSON representation includes augmented information in key names such as entities, attributes, relationships, and cardinalities. This additional context allows the LLM to make more accurate interpretations.

To verify the LLM's understanding of the JSON-represented diagrams, we use prompts such as "What are the entities?" and "What are the participating entities in the monitor relationship?". These prompts ensure that the LLM can accurately interpret and extract necessary details from the JSON format.

\subsection{Prompt Generation}
As mentioned earlier, we observe that asking LLMs to comment on the entire diagram reduces the quality and depth of the feedback, as LLM tends to focus on broad and high-level information such as the presence of entities and attributes. This approach overlooks more detailed feedback such as discussions on attribute types, cardinalities, or entity types.

To address this issue, we modified the AI system to provide feedback per relationship. This focused feedback ensures that the LLM can deliver more precise and detailed commentary on specific parts of the diagram. This method leverages the LLM's ability to handle smaller, more focused prompts, leading to more insightful and actionable feedback for students.

Moreover, this approach has shown to enhance the learning experience by allowing students to receive targeted guidance. They can iteratively refine their diagrams based on specific, actionable feedback, leading to a deeper understanding of the concepts and improved design skills. This methodology addresses the limitations of broader feedback and supports a more granular and effective learning process.

For this purpose, in addition to pruning the ERD around the selected relationship, we have to extract relevant requirements for each particular relationship. ensures that the feedback is contextual and directly applicable to the specific part of the diagram the student is working on.



\noindent\textbf{Requirements Matching}:
We find that utilizing LLMs to identify the relevant requirement items is more accurate than employing vector similarity search. LLMs excel in understanding and interpreting the context of requirements, enabling them to accurately match the specific needs of each relationship within a diagram. This contextual comprehension allows LLMs to notice the details and subtleties that vector similarity search often overlooks.

We create the following prompt to extract the relevant requirements for each pruned ERD. The prompt includes the entire set of requirements and the pruned ERD submitted by the student:
{\small
\begin{verbatim}
{
  "task": "Select the relevant items from 
    the problem statements for explaining 
    the given entity-relationships.",
  "entity-relationships": $submitted-erd,
  "problem-statements": "$problem-statments"},
  "relevant-statements": [ {
    "description": ..., 
    "rubrics":..., 
    "questions": ...}, ... ]
} 
\end{verbatim}
}

\noindent\textbf{Feedback Generation:}
After retrieving the relevant statements we construct the final prompt using the following template:
{\small
\begin{verbatim}
{
  "task": "Provide feedback based on the submitted
    relationship and participating entities, 
    and their attributes based on the provided 
    solution and problem statements. 
    The submission only contains one relationship.",
  "context": {
    "statement": "$relevant-statements",
    "submission": "$submitted-erd"
  },
  "output": {
    "feedback": "..."
  }
},
\end{verbatim}
}
\noindent where \$relevant-statements and \$submitted-erd are the JSONs for the selected relevant requirements and pruned ERD. 

\noindent\textbf{FAQ generation}:
We found that using a separate prompt for FAQs, which includes the previously generated feedback, results in more informative responses. We generate the FAQ using the following prompt:

{\small
\begin{verbatim}
{ 
  "task" : "What are potential follow up questions and
    answers regarding the provided feedback for the 
    submitted ERD. Answer the questions with detailed
    explanation.",
  "feedback": "$feedback",
  "problem-statements": "$relevant-statements",
  "submission": "$submitted-erd",
  "output": [ {{"question": ..., "answer": ...}}, ...]
},
\end{verbatim}
}
\noindent where \$feedback is the response from the previous prompt.

%% file: qualitative.tex
\section{Qualitative  Study}
In this section, we present our findings on the feedback generated by the LLM.

\noindent\textbf{Isolating a relationship}:
When asking the LLM to generate feedback on an entire diagram, the feedback often tends to be shallow and focuses on superficial features like attributes. However, when isolating a specific relationship, the feedback becomes more detailed and covers more in-depth aspects. This happens because the LLM has to manage a higher level of complexity and providing comprehensive feedback on an entire diagram can be cognitively demanding for the LLM (and for the students). To handle this, LLM may prioritize providing general observations that apply broadly across the diagram rather than discussing specifics that could overwhelm the initial feedback. By isolating a relationship, the scope of analysis is narrowed. This focused attention allows the LLM to allocate more resources to a smaller segment, leading to deeper and more detailed feedback on that specific part.\\

\noindent\textbf{Implicit Semantic Matching}:
When the submitted ERD uses different naming for entities, relationships, and attributes, the LLM can correctly relate them to the corresponding names in the problem statements and evaluate them accurately. In some cases, particularly for entity and relationship names, it suggests renaming to be consistent with the problem statements. For attributes, it mostly accepts the user's naming. For example, if the problem statement identifies total-amount as a derived attribute of the Invoice entity, but the submission names it total-due, the LLM confirms the correctness of the attribute type even though the naming is different.\\

\noindent\textbf{The importance of rubrics and questions}: 
We report the findings for the running example of the Patient-HasRecord-HealthRecord relationship. We assume that, as often happens, the students failed to identify HealthRecord as a weak entity. We compare the LLM's response when rubrics and questions are provided by the educator as part of the problem statements versus when the LLM does not have access to the rubrics. Without rubrics and questions, the LLM fails to generate correct feedback:
\begin{quote}
The submission correctly identifies the Patient and HealthRecord
entities along with their respective attributes. The Patient 
entity has id, name, address, and phone\_number as attributes,
with id being the candidate key. The HealthRecord entity has 
record\_id, disease, date, status, and description as attributes,
with record\_id being the key. The relationship HasRecord between
Patient and HealthRecord is also correctly identified with the 
correct cardinality. The Patient entity participates totally in 
the HasRecord relationship, which is correctly identified.
\end{quote}

\noindent By providing the rubric, the LLM's response becomes more accurate and reliable. The rubric offers clear guidelines and criteria for evaluation, enabling the LLM to generate feedback that closely aligns with the educator's expectations. This structured approach enhances the LLM's ability to identify key elements and common pitfalls in student responses, resulting in more precise and constructive feedback:
\begin{quote}

The submission correctly identifies Patient as a strong entity with attributes id, name, address, \\
and phone\_number. However, HealthRecord is incorrectly identified as having a key attribute record\_id. According to the problem statement, HealthRecord is a weak entity and should have a partial key instead of a candidate key. The relationship HasRecord between Patient and HealthRecord is correctly identified with Patient having a cardinality of 1 and HealthRecord having a cardinality of N. The total participation of HealthRecord in the relationship is also correctly identified.
    
\end{quote}

\noindent 
Without augmenting the problem statements with additional questions, the list of generated questions is less informative and has reduced educational value:

\begin{quote}
    What is the role of the 'id' attribute in the 'Patient' entity?
What is the purpose of the 'HasRecord' relationship in the ERD?
What does the 'status' attribute in the 'HealthRecord' entity indicate?
What is the significance of the 'total participation' type in the 'HasRecord' relationship?
\end{quote}

\noindent When augmenting the problem statements with the questions listed in Figure~\ref{fig:requirements}, the LLM has also added following question and answer:
\begin{quote}
Why is HealthRecord considered a weak entity in the ERD?
HealthRecord is considered a weak entity because it cannot exist independently without being associated with a Patient. It does not have a unique identifier that allows it to be uniquely identified among all other health records. Instead, it has a partial key, record\_id, which only uniquely identifies it among the health records of a specific patient.
\end{quote}

\noindent This question and answer are very informative, and having the questions guides the system to generate similarly informative questions and answers.

\noindent It is also worth noting that GPT-4's common knowledge is sufficient to answer the questions, for example:
\begin{quote}
    
Why must a weak entity have total participation in its identifying relationship?

A weak entity must have total participation in its identifying relationship because its existence depends on the strong entity. In this case, a HealthRecord cannot exist without being associated with a Patient. The total participation of HealthRecord in the HasRecord relationship signifies that every HealthRecord must be associated with a patient.
\end{quote} 

\noindent\textbf{Challenges in modeling superclass/subclass}:
We may decide to define a superclass for a group of entities via specialization. In these cases, we observe that the LLM cannot infer subclass attributes from the superclass. For example, if we model Nurse and Physician by factorizing their common attributes into Hospital-staff, the LLM cannot infer the subclass attributes from the superclass. This issue arises because, in our ERD representation, we used separate key-value pairs to represent specialization and union:
{\small
\begin{verbatim}
    "specializations": [{
      "name": "Hospital_staff",
      "subcategories": [
        {"name": "Nurse"},
        {"name": "Physician"} ],
      "type": ["disjoint"]
    }]
\end{verbatim}
}

\noindent Incorporating nested entities to represent subclass/superclass relationships in specialization and union may address this issue. We leave further investigation into these challenges for future work.\\

\noindent\textbf{Requirement granularity}:
To provide relationship-level feedback, it is important that each requirement item does not cover multiple relationships. For instance, the following requirement describes both the relationship between Invoice and Patient as well as the relationship between Invoice and Payable:
\begin{quote}
    The hospital issues invoices with unique account number and issue date for each patient. Invoice has start date and end date and includes all the payables in this duration. Payable items such as room charges, medication, and specific instructions provided by physicians. Each payable has a amount, date, description and a unique id. The hospital may issue multiple invoices for an account number with different issue dates.
\end{quote}
If we prune the ERD around the relationship between Payment and Invoice due to the mention of Payable in the requirement, the LLM expects to see that relationship in the pruned ERD, resulting in inaccurate feedback. The best way to address these issues is to modify the granularity of the requirements so that each requirement does not describe multiple relationships.\\

\noindent\textbf{Iterative refinement}: 
When a submission contains a considerable number of mistakes, the feedback provided by the LLM may not cover all errors in the initial review. This is because the LLM initially identifies and focuses on the most prominent or critical errors due to limitations in processing capacity. However, as students address some of the identified mistakes and refine their work, the LLM can recall and highlight additional errors.

This iterative refinement process not only aids in gradually improving the diagram but also helps students develop a deeper understanding of the conceptual design and the associated thought process. By systematically working through the mistakes, students can enhance their problem-solving skills and gain confidence in their ability to create accurate and effective diagrams. Additionally, this process mimics a natural learning progression, where addressing foundational issues first paves the way for recognizing and correcting more subtle errors, ultimately leading to a more thorough understanding and improved final output.

%% file: study.tex
\section{Pilot Study}
The system was piloted in a Database Systems course for the final project of the course. The course has an enrollment of 60 students, and students work in groups of at most five on their final project.\footnote{Our study has been approved by our institutional review board.}

The system has generated 242 pieces of feedback for the students, the teaching staff has reviewed all the AI feedback and intervened on 79 feedback by creating a discussion either because the AI feedback is unclear or inaccurate, or because students have further questions. The augmented discussion capabilities address concerns about relying on AI responses. 

\subsection{Student Self-Evaluation}
We use exit surveys for students to report a self-evaluation of the AI system, including the quality of feedback and other features such as FAQ and discussion threads. 84\% of the participants find the feedback helpful in improving their ERDs, and 80.0\% of the students rated the quality of feedback as 4 or 5 on a 5 scale rating with 5 being the highest. 

\begin{figure}
    \centering
    \includegraphics[width=0.9\linewidth]{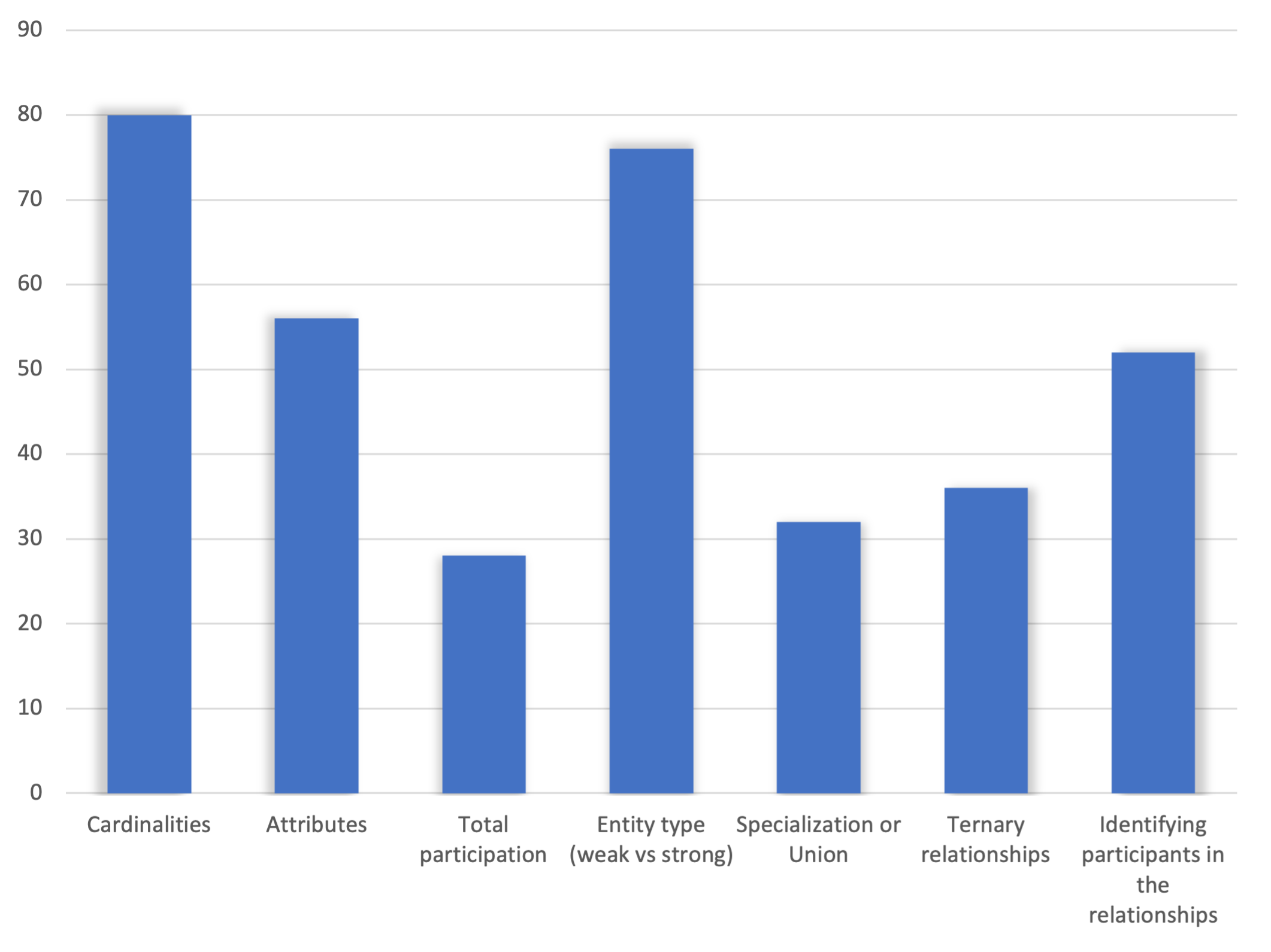}
    \caption{The concepts that students receive a feedback that helped them improve their ERD (from student perspective).}
    \label{fig:feedback_type_survey}
\end{figure}

The participants were asked to identify in which categories AI helped them improve their ERDs. The categories include Cardinalities, Identifying attributes, Total participation, Entity type, Specialization or Union, Ternary relationships,
and Identifying participants in the relationships. The outcome is shown in Figure~\ref{fig:feedback_type_survey}. Note that the feedback may not always be correct. However, based on previous offerings of the course, incorrect cardinalities and missing weak entities are the most common mistakes among students.

We also survey the students on FAQ and discussion component of the systems. 28\% strongly agree that the provided questions and answers are helpful, while 48.0\% agree, 20.0\% neutral, and 4\% disagree.
Moreover, most of the students find discussion threads helpful for addressing problems with LLM feedback (28\% strongly agree and 56\% agree).

\subsection{Expert Evaluation}
We manually evaluate the correctness of LLM feedback for each of the categories in the previous section. In addition to those categories, we also include identifying key attributes, relationship types (identifying vs non-identifying), and invalid relationships. 
\noindent\textbf{Evaluation metric}:
For each category, we define a positive as the presence of a mistake in the student submission and a negative as a correct submission. We define true positive (TP) as the LLM correctly detecting a mistake, false positive (FP) as the LLM incorrectly recognizing a mistake, true negative (TN) as the LLM correctly confirming the correctness of the submission, and false negative (FN) as the LLM incorrectly confirming the correctness of an incorrect submission.
We report precision: $\frac{TP}{FP+TP}$, recall: $\frac{TP}{TP+FN}$ and $F_1$ score: $\frac{2*\text{precision}*\text{recall}}{\text{precision}+\text{recall}}$ as the harmonic mean of precision and recall. These metrics are commonly used in evaluating machine learning methods.

\begin{table}[]
    \centering
{\small
    \caption{Comparison of feedback performance for different types of mistakes.}
    \begin{tabular}{l|c c c}
    \toprule
         & Precision & Recall & $F_1$ Score \\
         \midrule
    Relationship Participants     &  0.83     &   0.91  &  0.87   \\
    Cardinalities     &  0.96    &  0.93    &   0.94  \\
    Attributes     &   0.95   &   0.77  &   0.85  \\
    Attribute Types     &   0.93   &  0.87   &   0.90  \\
    Keys     &   1.0   &   0.45   &  0.62   \\
    Ternary Relationships     &   1.0   &  0.91   &   0.95  \\
    Total Participation    &  1.0    &  0.2   &   0.33  \\
    Relationship Types     &   1.0   &   0.57  &   0.73  \\
    Specialization or Union     &  0.29    &   0.50  &  0.36   \\
    Entity Types     &  0.78    &   1.0  &  0.88   \\
    Invalid Relationships     &  0.88    & 0.88    & 0.88    \\
    \bottomrule
    \end{tabular}
    }
    \label{tab:evaluation}
\end{table}

\noindent\textbf{Evaluation results}: Table~1 shows the results of evaluating 100 LLM feedbacks. For most categories, the LLM has a high precision, indicating that when it reports a mistake, it is mostly correct. However, for most categories, it has a lower recall compared to precision, meaning it may overlook some mistakes in the submissions. In other words, it does not recall all the mistakes. This partly happens because the LLM feedback may partially address the mistakes, and as students refine their diagrams, more mistakes are uncovered. Among these categories, total participation has the lowest recall. In the study environment, there is only one total participation accompanied by a weak entity, and in most cases, the LLM focuses on the weak entity type rather than total participation.

Specialization or Union has the lowest precision among these categories. As discussed earlier, the LLM cannot correctly detect subclass and superclass relationships due to the used EERD representation, resulting in correct designs being reported as mistakes. Although the LLM feedback mostly concerns missing attributes of subclasses, we reported it in the Specialization or Union category as it pertains to this concept rather than attributes.

Comparing the $F_1$ score and student self-reports (Figure~\ref{fig:feedback_type_survey}) on the overlapping categories shows interesting correlations regarding how LLM feedback helps students improve their diagrams. Categories with higher $F_1$ scores, such as Cardinalities, Entity Type, Relationship Participants, and Attributes, were more helpful for students in improving their diagrams. On the other hand, categories with the lowest scores, such as Total Participation and Specialization or Union, were less helpful.

%% file: discussion.tex
\section{Conclusion}
In this work, we introduce a new LLM-based feedback generation tool to enhance conceptual design learning. Our tool accepts a new environment via a set of requirements, and for each requirement item, the educator can design a specific natural language rubric and important questions. We use the requirements, provided rubrics, and questions to contextualize LLMs in generating feedback for student ERD designs. Our qualitative and quantitative evaluation of the system confirms its usefulness in providing informative feedback to students. This outcome has also been confirmed with student self-evaluation reports. Improving the quality of LLM feedback generation can help bridge the gap between available teaching resources and growing student enrollment in high-demand programs such as Computer Science.
